\documentclass[amsmath,amssymb,twocolumn,nofootinbib]{revtex4-1}
\usepackage{graphicx}
\usepackage{epsfig}
\usepackage{hyperref}
\usepackage{color}
\usepackage{soul}

\begin{document}
\title{Future soft singularities, Born-Infeld-like fields and particles}
\author{Olesya Galkina}
\email{olesya.galkina@cosmo-ufes.org}
\affiliation{PPGFis, CCE - Universidade Federal do Esp\'{i}rito Santo, zip 29075-910, Vit\'{o}ria, ES, Brazil\\
Dipartimento di Fisica e Astronomia, Universit\`a di Bologna\\ and INFN,  Via Irnerio 46, 40126 Bologna,
Italy}
\author{Alexander~Yu.~Kamenshchik}
\email{kamenshchik@bo.infn.it}
\affiliation{Dipartimento di Fisica e Astronomia, Universit\`a di Bologna\\ and INFN,  Via Irnerio 46, 40126 Bologna,
Italy,\\
L.D. Landau Institute for Theoretical Physics of the Russian
Academy of Sciences,\\
 Kosygin street. 2, 119334 Moscow, Russia}
\begin{abstract}
We consider different scenarios of the evolution of the universe, where the singularities or some non-analyticities in the geometry of the spacetime are present, trying to answer the following question: is it possible to conserve some kind of notion of particle corresponding to a chosen quantum field present in the universe when the latter approaches the singularity? We study scalar fields with different types of Lagrangians, writing down the second-order differential equations for the linear perturbations of these fields in the vicinity of a singularity. If both independent solutions are regular, we construct the vacuum state for quantum particles as a Gaussian function of the corresponding variable. If at least one of two independent solutions has a singular asymptotic behavior, then we cannot define the creation and the annihilation operators and construct the vacuum. This means that the very notion of particle loses sense. We show that at the approaching to the Big Rip singularity, particles corresponding to the phantom scalar field driving the evolution of the universe must vanish, while particles of other fields still can be defined. In the case of the model of the universe  described by the tachyon field with a special trigonometric potential, where the Big Brake singularity occurs, we see that the (pseudo) tachyon particles do not pass through this singularity. Adding to this model some quantity of dust, we slightly change the characteristics of   this singularity and tachyon particles survive. Finally, we consider a model with the scalar field with the cusped potential, where the phantom divide line crossing occurs. Here the particles are well defined in the vicinity of this crossing point.   
\end{abstract}
\maketitle

\section{Introduction}
The problem of cosmological singularities has been attracting the attention
of theoreticians working in gravity and cosmology at least since the early fifties. 
In the sixties general theorems about the
conditions for the appearance of singularities were proven \cite{Hawk,Pen}
and the oscillatory regime of approaching the singularity \cite{BKL},  called also
``Mixmaster universe'' \cite{Misner}  was discovered. 
Intuitively, when one hears the word ``cosmological singularity'' one thinks about 
a universe with a vanishing cosmological radius, i.e. about the Big Bang and the Big Crunch 
singularities.

Basically, until the end of
nineties almost all discussions about singularities were devoted to the Big
Bang and the Big Crunch singularities, which are characterized by a vanishing
cosmological radius.
However, kinematical investigations of Friedmann cosmologies have raised the 
question about the
possibility of a sudden future singularity occurrence \cite{sudden1}, 
characterized by a diverging $\ddot{a}$ whereas both the scale
factor $a$ and $\dot{a}$ are finite. Then the Hubble parameter $H=\dot{a}
/a~ $and the energy density $\rho $ are also finite, while the first
derivative of the Hubble parameter and the pressure $p$ diverge. 
Until recent years, however, the sudden future singularities attracted rather a limited interest 
of researchers. The situation has changed  in the new millennium, when  plenty of publications 
devoted to such singularities have appeared \cite{sudden2}--\cite{Nojiri:2004pf}.

In the investigations devoted to sudden singularities one can distinguish three main topics. The first of them deals with the question of  the compatibility of the models possessing soft singularities with observational data \cite{sudden6,Mar,tach1,Tomasz}.
The second direction is connected with the study of quantum effects \cite{Shtanov1,we-tach,sudden8,Haro,Kiefer4,quantum,Kiefer2,quantum1,Claus-we,Barbaoza,Calderon}. Here one can 
see two subdirections: the study of quantum corrections to the effective Friedmann equation, which can eliminate classical singularities or, at least, change their form \cite{Shtanov,sudden8,Haro}, and the study of solutions of the Wheeler-DeWitt equation for the quantum state of the universe in the presence of sudden singularities \cite{Kiefer4,quantum,Kiefer2,quantum1,Barbaoza}. 
The third direction is connected with the opportunity of the crossing of sudden singularities in classical cosmology 
\cite{Lazkoz,Lazkoz3,Lazkoz1,tach2,quantum1}.

A particular feature of the sudden future singularities is their softness 
\cite{Lazkoz}. As the Christoffel symbols depend only on the first
derivative of the scale factor, they are regular at these singularities.
Hence, the geodesics have a good behavior and they can cross the singularity 
\cite{Lazkoz}. One can argue that the particles crossing the singularity
will generate the geometry of the spacetime, providing in such a way a soft rebirth  of the universe after the
singularity crossing \cite{tach2}. Note that the opportunity of crossing of some kind of cosmological 
singularities were noticed already in the early paper by Tipler \cite{Tipler}.
A rather close idea of integrable singularities in black holes, which can give origin to a cosmogenesis, 
was recently put forward in \cite{Lukash,Lukash1}.

Another remarkable feature of the soft future singularities is their capacity to induce changes in the equations of state of the matter present in a universe under consideration. Moreover, the form of the matter Lagrangian can also be changed. These effects were considered in  \cite{Paradox1,myreview,myreview1}. 
The effects of the matter transformation occur sometimes also without singularities, but only in the presence of some non-analyticities in the geometry of the spacetime
\cite{cusped,cusped1}. These phenomena have also some kinship with those of the singularity crossing \cite{myreview,myreview1}. 

While the crossing of the future sudden singularities does not look too counterintuitive, it is more difficult to imagine the crossing of the Big Bang -- Big Crunch type singularities. However, already in this millennium different approaches to this problem were developed. First of all, let us mention the ekpyrotic scenario \cite{Turok,Turok1,Lehners2}. One of the features of this scenario is 
usage of models with the presence of two scalar fields. Another series of works, partially connected with this scenario \cite{Bars,Bars1,Bars2,Bars3,Bars4}
was explicitly devoted to the detailed description of crossing of the Big Bang - Big Crunch singularity. There the presence of two scalar fields and the Weyl invariance have an essential role. A similar transitions through the Big Bang - Big Crunch singularities in the model with one scalar field were considered in papers \cite{we-Bang,we-Bang1,we-Bang2,we-Bang3}, where the transitions between the Jordan frames and the Einstein frames and ideas of the analytic continuation were used.  The transformations between the frames as a tool describing the origin of the universe and the crossing of singularities were used also in papers 
\cite{Wetterich,Wetterich1,Wetterich2}. 
  
  The development of the modern theoretical physics, including cosmology,  can convince us that the quantum theory is more fundamental than classical and that the classicality can arise as a temporary phenomenon. At the level of simple toy models it was illustrated, for example, in paper \cite{we-basis}.  Thus, it is very logical to suggest that the consistent description of the singularity crossing can be achieved in quantum theory as was done in paper \cite{Lehners1}. 

While the full theoretical description of the process of the crossing of the cosmological singularities is possible in the framework of the complete quantum theory of gravity, the application of the methods of the quantum field theory on curved classical background 
\cite{DeWitt, Birrell, Parker} can also bring some interesting results.
In the present paper we shall use these methods to study another aspect of the presence of soft singularities and non-analyticities of geometry - we are interested in the behavior of quantum particles in the vicinity of these particular spacetime points.

It is well known that the very notion of particle becomes complicated when one considers the quantum field theory on a curved spacetime background \cite{DeWitt, Birrell, Parker}. Let us recapitulate the general procedure for the definition of the particles on the example of a scalar field filling a flat
Friedmann universe  with the metric 
\begin{equation}
ds^2=dt^2-a^2(t)dl^2.
\label{Fried}
\end{equation}
The Klein-Gordon equation for the minimally coupled scalar field $\phi$ with the potential $V(\phi)$ is 
\begin{equation}
\Box\phi +V'(\phi) = 0,
\label{K-G}
\end{equation}
where $\Box$ is the d'Alambertian. One can consider a spatially homogeneous solution of this equation $\phi_0$, depending only on time $t$ as a classical background.
A small deviation from this background solution can be represented as a sum of  Fourier harmonics satisfying linearized equations
\begin{equation}
\ddot{\phi}(\vec{k},t)+3\frac{\dot{a}}{a}\dot{\phi}(\vec{k},t)+\frac{\vec{k}^2}{a^2}\phi(\vec{k},(t)) + V''(\phi_0(t))\phi(\vec{k},(t)) = 0.
\label{K-G1}
\end{equation}
The corresponding quantized field is represented in the following form
\begin{equation}
\hat{\phi}(\vec{x},t) = \int d^3\vec{k}(\hat{a}(\vec{k})u(k,t)e^{i\vec{k}\cdot\vec{x}}+\hat{a}^+(\vec{k})u^*(k,t)e^{-i\vec{k}\cdot\vec{x}}),
\label{quant}
\end{equation}  
where the creation and the annihilation operators satisfy the standard commutation relations:
\begin{equation}
[\hat{a}(\vec{k}), \hat{a}^+(\vec{k}')]=\delta(\vec{k}-\vec{k}'),
\label{quant1}
\end{equation}
while the basis functions $u$ satisfy the linearized equation (\ref{K-G1}).
These basis functions should be normalized so that the canonical commutation relations between the field $\phi$ and its canonically conjugate momentum 
$\hat{{\cal P}}$ were satisfied
\begin{equation}
[\hat{\phi}(\vec{x},t), \hat{{\cal P}}(\vec{y},t')]=i\delta(\vec{x}-\vec{y}).
\label{canon}
\end{equation}
Taking into account the fact  that for the minimally coupled scalar field the momentum is 
\begin{equation}
\hat{\cal P}(\vec{x},t)=a^3\dot{\phi}(\vec{x},t)
\label{canon1}
\end{equation}
the commutation relation (\ref{quant1}) and the Fourier representation for the  Dirac delta function, one easily shows that 
the relation (\ref{canon}) is satisfied if 
\begin{equation}
u(k,t)\dot{u}^*(k,t)-u^*(k,t)\dot{u}(k,t)=\frac{i}{(2\pi)^3a^3(t)}.
\label{canon2}
\end{equation}
The linearized equation (\ref{K-G1}) has two independent solutions. As for functions $u$, one can take different linear combinations of these solutions chosen in such a manner that the Wronskian relation (\ref{canon2}) is satisfied. Different choices of these functions determine different choices of the creation and the annihilation operators and different vacuum states on which the Fock spaces can be constructed. In the Minkowski spacetime a preferable choice  simply  corresponds to the plane waves. In the de Sitter spacetime it is common to define the Bunch-Davies vacuum \cite{Bunch}, which in the limit of large wave numbers is close to the Minkowski vacuum.  In any case, in order to have some definition of particle it is necessary to obtain two independent non-singular solutions of Eq. (\ref{K-G1}). However, it is a non-trivial requirement in the situations when a singularity or other kind of irregularity of the spacetime geometry occurs. One can easily understand that this  is connected with the presence of the time-dependent scale factor $a(t)$ in the right-hand side of the relation (\ref{canon2}). Let us mention that the second-order differential equation for the perturbations on the highly non-trivial background was studied in paper \cite{Lehners}.

It is convenient also to construct explicitly the vacuum state for quantum particles as a Gaussian function of the corresponding variable. Let is introduce an operator
\begin{equation}
\hat{f}(\vec{k},t) = (2\pi)^3(\hat{a}(\vec{k})u(k,t)+\hat{a}^+(-\vec{k})u^*(k,t)).
\label{vac-exp}
\end{equation}
Its canonically conjugate momentum is 
\begin{equation}
\hat{p}(\vec{k},t) = a^3(t)(2\pi)^3(\hat{a}(\vec{k})\dot{u}(k,t)+\hat{a}^+(-\vec{k})\dot{u}^*(k,t)).
\label{vac-exp1}
\end{equation}
Now we can express the annihilation operator as 
\begin{equation}
\hat{a}(\vec{k}) =i\hat{p}(\vec{k},t)u^*(k,t)-ia^3(t)\hat{f}(\vec{k},t)\dot{u}^*(k,t),
\label{vac-exp2}
\end{equation}
where we have used the  Wronskian relation (\ref{canon2}).
Representing the operators $\hat{f}$ and $\hat{p}$ as 
\begin{equation}
\hat{f} \rightarrow f,\ \ \hat{p} \rightarrow -i\frac{d}{df},
\label{vac-exp3}
\end{equation}
one can write down the equation for the corresponding vacuum state in the following form:
\begin{equation}
\left(u^*\frac{d}{df}-ia^3\dot{u}^*f\right)\Psi_0(f) = 0.
\label{vac-exp4}
\end{equation}
The normalized solution to Eq. (\ref{vac-exp4}) is (up to a non-essential constant) 
\begin{equation}
\Psi_0(f) = \frac{1}{\sqrt{|u(k,t)|}}\exp\left(\frac{ia^3(t)\dot{u}^*(k,t)f^2}{2u^*(k,t)}\right).
\label{vac-exp5}
\end{equation}

As we have already mentioned, in the present paper we study what happens with quantum fields in curved spacetimes in the vicinity of  singularities or non-analyticities and analyze when the regular solutions of the corresponding linearized equations exist. 
Besides, we shall check if it possible to construct the vacuum states which look like Eq. (\ref{vac-exp5}). 
The structure of the paper is the following: in the second section we consider the traditional Big Bang - Big Crunch and Big Rip singularities. The third section is devoted to some models based on tachyon fields, revealing the Big Brake and other soft future singularities and the effects of transformations of  matter fields \cite{we-tach}. In the fourth section we consider a particular cosmological model \cite{cusped,cusped1} describing the smooth transformation between the standard and phantom scalar fields. The final section includes some conclusive remarks. 

\section{Big Bang -- Big Crunch, Big Rip and particles}

At the  Big Bang or the Big Crunch singularity a universe has a vanishing volume or in the case of homogeneous and isotropic Friedmann universe, which we consider in this paper, 
the vanishing scale factor $a$. This means that the Wronskian, which is inversely proportional to $a^3$ (see Eq.  (\ref{canon2})), becomes singular. This points out that it could be impossible to construct the non-singular basis functions in the vicinity of the singularity, and, correspondingly, one cannot introduce a Fock vacuum and the operators of creation and annihilation. To confirm this statement let us consider a simple case of a flat Friedmann universe filled with a perfect fluid with the equation of state 
\begin{equation}
p = w\rho,
\label{eq-of-state}
\end{equation}
where $p$ is the pressure, $\rho$ is the energy density and $w$ is a constant such that $-\frac13 < w \leq 1$. 
The law of expansion of the universe is
\begin{equation}
a(t) = a_0t^{\frac{2}{3(1+w)}}. 
\label{expansion}
\end{equation}
We can consider, for example, a free massive scalar field living in this universe. Then Eq. (\ref{K-G1}) looks as 
\begin{equation}
\ddot{u}(\vec{k},t)+\frac{2}{(1+w)t}\dot{u}(\vec{k},t)+\frac{k^2}{a_0^2t^{\frac{4}{3(1+w)}}}u(\vec{k},t)+m^2u(\vec{k},t)=0.
\label{bang}
\end{equation}
Obviously, considering Eq. (\ref{bang}) at $t \rightarrow 0$, we can neglect the massive term with respect to the term inversely proportional to $t^{\frac{4}{3(1+w)}}$.  
After this it is easy to find that
\begin{eqnarray}
&&u(\vec{k},t) = c_1t^{\frac{w-1}{2(1+w)}}J_{\frac{1-w}{2(1+w)}}\left(\frac{3k(1+w)}{a_0(1+3w)}t^{\frac{(1-w)(1+3w)}{(1+3w)^2}}\right)
\nonumber \\
&&+c_2t^{\frac{w-1}{2(1+w)}}Y_{\frac{1-w}{2(1+w)}}\left(\frac{3k(1+w)}{a_0(1+3w)}t^{\frac{(1-w)(1+3w)}{(1+3w)^2}}\right).
\label{bang1}
\end{eqnarray} 
Here, $J$ and $Y$ are the corresponding Bessel functions. We see that the term, proportional to the function $Y$ becomes singular when $t \rightarrow 0$ and, hence, we
do not have two independent non-singular solutions for the basis functions and cannot construct the vacuum and the Fock space.
Note, that this conclusion is valid even if for the model under consideration one manages to describe the Big Bang - Big Crunch 
singularity crossing, using some of the approaches mentioned in the Introduction.

Now, let us consider an extreme opposite case - the Big Rip singularity \cite{Star-Rip,Rip,phantom}. The simplest model, where this singularity arises, is the Friedmann universe 
filled with a perfect fluid with a constant equation of state parameter $w$ such that $w < -1$. In this case the scale factor behaves as 
\begin{equation}
a(t) = a_0(-t)^{\frac{2}{3(1+w)}},
\label{Rip}
\end{equation}
 and when $t \rightarrow 0_-$ the scale factor tends to $\infty$. The equation for the perturbations of the massive scalar field on this background have the same form 
 as Eq. (\ref{bang}), but now we can neglect the term $\frac{k^2}{a_0^2t^{\frac{4}{3(1+w)}}}u(\vec{k},t)$, which tends to zero as $t \rightarrow 0_-$. Thus, the solution of the corresponding equation is 
\begin{eqnarray}
&&u(\vec{k},t) = c_1(-t)^{\frac{w-1}{2(1+w)}}J_{\frac{w-1}{2(1+w)}}\left(-mt\right)
\nonumber \\
&&+c_2(-t)^{\frac{w-1}{2(1+w)}}Y_{\frac{w-1}{2(1+w)}}\left(-mt\right).
\label{Rip1}
\end{eqnarray} 
Both independent solutions are now regular at $t \rightarrow 0_-$ and we can construct the Fock vacuum. Thus, nothing special happens with particles when universe 
enters into the Big Rip singularity. Let us construct this vacuum state in the vicinity of the singularity explicitly, using the formula
\eqref{vac-exp5}. In the vicinity of the Big Rip we can write down the basis function using the independent solutions \eqref{Rip1}
and keeping only the leading terms as follows:
\begin{equation}
u(\vec{k},t) = A + iB(-t)^{\frac{w-1}{1+w}}.
\label{Rip10}
\end{equation}
This function should satisfy the Wronskian relation \eqref{canon2}, with the scale factor given by the formula \eqref{Rip}.
 It means that the constants $A$ and $B$ satisfy the equation
 \begin{equation}
 AB = \frac{(1+w)}{(2\pi)^3a_0^3(w-1)}.
 \label{Rip11}
 \end{equation}
Then,
\begin{equation}
\Psi_0(f) \sim \frac{1}{A}\exp\left(-\frac{1}{(2\pi)^3A^2}f^2\right).
\label{Rip12}
\end{equation}
The Gaussian exponent is well defined in the vicinity of the Big Rip singularity. We still have the freedom to choose the value of the positive constant $A$. We know, for example, that in the case of the de Sitter spacetime, one can fix an analogous freedom by requiring that the vacuum has a standard Minkowski form in the infinitely remote past. Here, we cannot follow the evolution of our basis function to the past infinity and, thus, we leave the value of the constant $A$ unspecified.
However, for any choice of this constant, the function \eqref{Rip12} has a regular behavior. Let us note that at least up to our knowledge there are no attempts to describe the Big Rip singularity crossing. Thus, the regular behavior of the quantum particles approaching the Big Rip singularity does not mean that such a singularity can be crossed, or they can survive such a crossing. Nevertheless, the fact of the regular behavior of functions, entering into the formulas \eqref{Rip1} and \eqref{Rip12} looks interesting.

Let us consider a slightly more complicated situation when the evolution of type (\ref{Rip}) is provided by the presence of the phantom scalar field with the negative kinetic term and an exponential potential:
\begin{equation}
L = -\frac12g^{\mu\nu}\phi_{,\mu}\phi_{,\nu} - V_0\exp(-\alpha\phi).
 \label{Rip2}
 \end{equation}
 The Friedmann equation is now 
 \begin{equation}
 \frac{\dot{a}^2}{a^2}=-\frac12\dot{\phi}^2+V_0\exp(-\alpha\phi),
 \label{Rip3}
 \end{equation}
 while the Klein-Gordon equation is 
 \begin{equation}
 \Box\phi +\alpha V_0\exp(-\alpha\phi) = 0.
 \label{Rip4}
 \end{equation}
 If we choose 
 \begin{equation}
 V_0=\frac{2(1-w)}{9(1+w)^2}
 \label{Rip5}
 \end{equation}
 and 
 \begin{equation}
 \alpha = 3\sqrt{-(1+w)},
 \label{Rip50}
 \end{equation}
 then we have the evolution (\ref{Rip}) and the background solution for the phantom scalar field is 
 \begin{equation}
 \phi(t) = \frac{2}{3\sqrt{-(1+w)}}\ln(-t).
 \label{Rip6}
 \end{equation}
 Before writing down the equation for the linear perturbations we should substitute into the Klein-Gordon equation \eqref{Rip4} the expression for 
 $\frac{\dot{a}}{a}$ following from the Friedmann equation \eqref{Rip3}. Then we have the equation which includes only the scalar field and its derivatives. 
 The equation for the linear perturbations is now 
\begin{eqnarray}
&&\ddot{u}(\vec{k},t)+\frac{1-w}{(1+w)t}\dot{u}(\vec{k},t)+\frac{k^2}{a_0^2t^{\frac{4}{3(1+w)}}}u(\vec{k},t)\nonumber \\
&&+\frac{1-w}{(1+w)t^2}u(\vec{k},t)=0.
\label{Rip7}
\end{eqnarray}
 In the vicinity of the Big Rip singularity $t \rightarrow 0_-$, the solution of Eq. (\ref{Rip7}) behaves as 
 \begin{equation}
 u(\vec{k},t) = c_1 (-t)^{\kappa_1}+c_2(-t)^{\kappa_2},
 \label{Rip8}
 \end{equation}
 where 
 \begin{equation}
 \kappa_1 = \frac{w}{1+w} + \sqrt{\frac{2w^2-1}{(1+w)^2}} > 0,
 \label{Rip9}
 \end{equation}
 \begin{equation}
 \kappa_2 =  \frac{w}{1+w} - \sqrt{\frac{2w^2-1}{(1+w)^2}} < 0.
 \label{Rip10}
 \end{equation}
 Thus, the second  solution in (\ref{Rip8}) is singular as  $t \rightarrow 0_-$ and we cannot construct the Fock space for it.
 
\section{Tachyon model and soft singularities}

The discovery of cosmic acceleration \cite{cosmic} stimulated searches of the so-called dark energy responsible for this effect
\cite{dark,dark1}. One of the possible candidates for this role was tachyon field, arising  in  string theories \cite{Sen,Feinstein,Padman,Kofman}. 
As a  matter of fact what is called tachyon field is a modification of an old idea of Born and Infeld \cite{B-I}, that the kinetic term of a field can have a non-polynomial form.    
The Lagrangian of the tachyon field $T$ has the form
\begin{equation}
L = -V(T)\sqrt{1-g^{\mu\nu}T_{,\mu}T_{,\nu}},
\label{tach}
\end{equation}
which for a spatially homogeneous field becomes 
\begin{equation}
L=-V(T)\sqrt{1-\dot{T}^{2}}.
\label{Lagrangian-T}
\end{equation}  
The energy density corresponding to (\ref{Lagrangian-T}) is 
\begin{equation}
\rho = \frac{V(T)}{\sqrt{1-\dot{T}^2}},
\label{en-den}
\end{equation}
while the pressure is negative and equal to 
\begin{equation}
p=-V(T)\sqrt{1-\dot{T}^{2}}.
\label{Lagrangian-T1}
\end{equation}
The negativity of the pressure makes the tachyon field a good candidate for the dark energy role. 
The field equation for the tachyon field is 
\begin{equation}
\frac{\ddot{T}}{1-\dot{T}^2} + 3H\dot{T} +\frac{V_{,T}}{V(T)} = 0.
\label{KGT}
\end{equation}
There is also a great freedom for the choice of the potential $V(T)$. 
In the paper \cite{we-tach} a very particular potential, depending on the trigonometrical functions was chosen:
\begin{equation}
V\left(T\right)=\frac{\Lambda\sqrt{1-\left(1+w\right)\cos^{2}\left(\frac{3}{2}\sqrt{\Lambda\left(1+w\right)}\right)T}}{\sin^{2}\left[\frac{3}{2}\sqrt{\Lambda(1+w)}T\right]},
\label{poten}
\end{equation} 
where $\Lambda$ is a positive constant and $ -1 < w \leq 1$.
What is the origin of this potential? If one consider a flat Friedmann model filled with the cosmological constant $\Lambda$ and a perfect fluid with a constant barotropic index 
$w$ then one can find an exact solution for the cosmological evolution. Then it is possible to reconstruct the potential $V(T)$ of the tachyon field generating this exact solution as a particular solution of the system which includes the Friedmann equation and Eq. (\ref{KGT}). This potential is nothing but the potential (\ref{poten})  from the paper \cite{we-tach}. However, the dynamics of the Friedmann model based on the tachyon field with the potential (\ref{poten}) is more rich than that of the model with two fluids, because 
the model with tachyon has more degrees of freedom. The case when the parameter $w$ is positive is particularly interesting. To study this case it is convenient to rewrite the Klein-Gordon-type equation (\ref{KGT}) as  a dynamical system of two  first-order differential equations:
\begin{align}
\dot{T} & =s,\\
\dot{s} & =-3\sqrt{V}\left(1-s^{2}\right)^{\frac{3}{4}}s-\left(1-s^{2}\right)\frac{V_{,T}}{V}.
\label{dyn-sys}
\end{align}   
The phase portrait for this dynamical system is presented on the figure below, which was taken from the paper \cite{we-tach}.  One can see that the potential (\ref{poten}) is well defined inside the rectangle, where 
$-1 \leq s \leq 1$ and $T_3 \leq T \leq T_4$, with 
\begin{equation}
T_3 = \frac{2}{3\sqrt{(1+w)\Lambda}} {\rm arccos} \frac{1}{\sqrt{1+w}},
\label{T3}
\end{equation} 
\begin{equation}
T_4 = \frac{2}{3\sqrt{(1+w)\Lambda}} \left(\pi - {\rm arccos} \frac{1}{\sqrt{1+w}}\right).
\label{T4}
\end{equation}
The analysis of this dynamical system shows that there are two families of the trajectories, one of them tends to the center of the rectangle, where $s = 0$ and 
$T = \frac{\pi}{3\sqrt{(\Lambda(1+w)}}$. Such a cosmological evolution is very close to one in the standard $\Lambda$CDM model. Another family  includes the trajectories  which tend  to corners of our rectangle: one with $s=-1$ and $T=T_3$ and the symmetric one with $s=1$ and $T=T_4$. \\
\vspace{0.5cm}
\begin{figure}[h]
\includegraphics[scale=0.5]{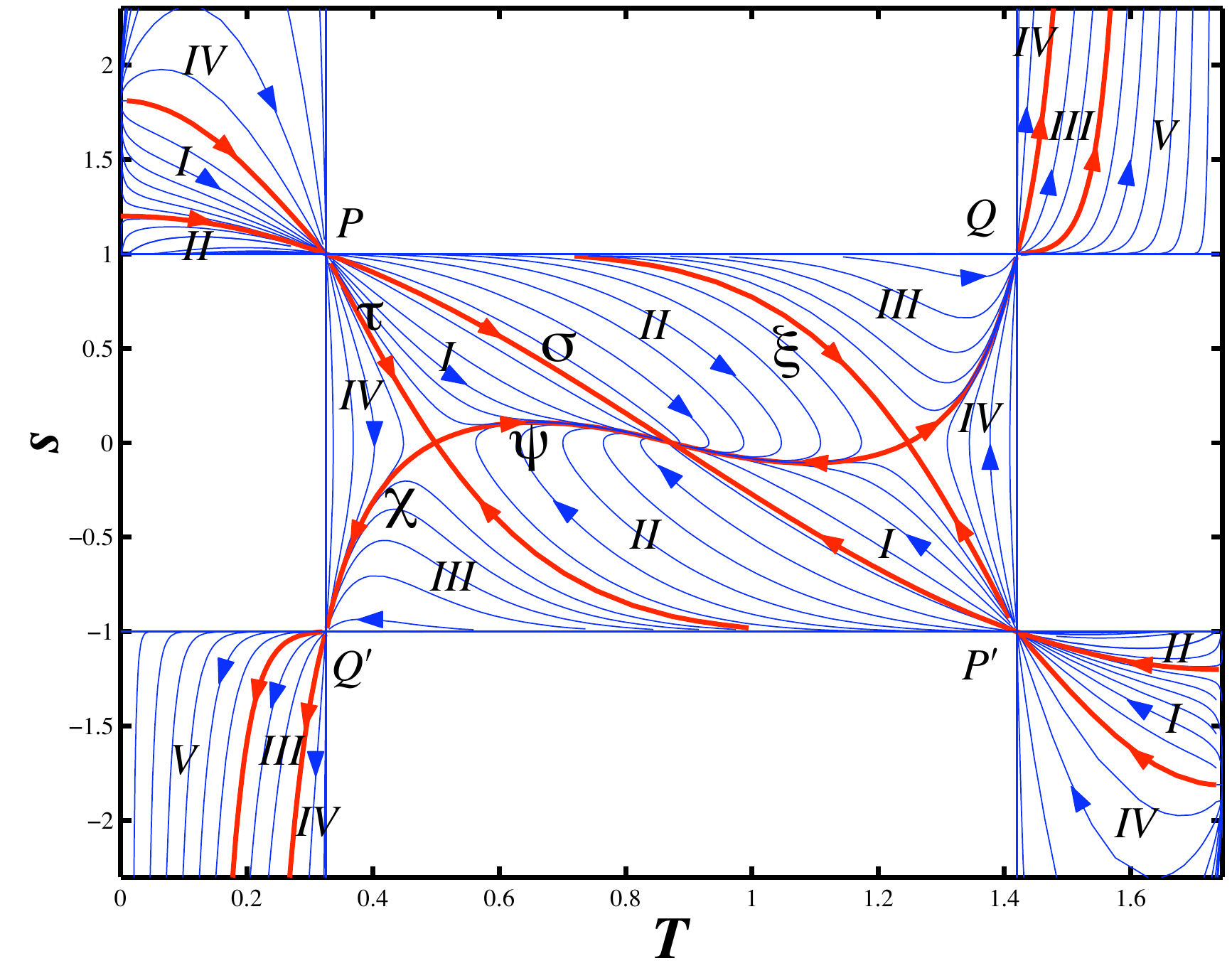}
{\small
Phase portrait of the model for a positive \textcolor{blue}{$w$}}.
\end{figure}
What happens with the universe approaching, for example, the lower left corner? The expression under the square root in the potential (\ref{poten}) tends to zero and the kinetic expression  
$\sqrt{1-s^2}$ tends to zero and it looks like we cannot cross the corner. At the same time it is easy to see  there is no cosmological singularity here. Moreover, the differential equations are also regular. In paper \cite{we-tach}  the only possible way out was suggested. The Lagrangian changes its form in such a way 
that the equations of motion conserve their form. The new Lagrangian is 
\begin{equation}
L=W(T)\sqrt{\dot{T}^2-1},
\label{pseudo}
\end{equation}
where 
\begin{equation}
W(T)=\frac{\Lambda\sqrt{\left(1+w\right)\cos^{2}\left(\frac{3}{2}\sqrt{\Lambda\left(1+w\right)}\right)T-1}}{\sin^{2}\left[\frac{3}{2}\sqrt{\Lambda(1+w)}T\right]},
\label{pseudo-poten}
\end{equation}
and the new field (or a new form of the old field) is called pseudotachyon \cite{we-tach}. This field arises when the universe enters into the left lower infinite strip on the figure.
Note that the Friedmann equation for the universe filled with the pseudotachyon field is 
\begin{equation}
\frac{\dot{a}^2}{a^2} = \frac{W(T)}{\sqrt{\dot{T}^2-1}}.
\label{pseudo-Fried}
\end{equation}

 Let us describe in detail what happens with the field  when it crosses  the corner.
The spatially homogeneous part of the field $T$ behaves as 
\begin{equation} 
T = T_3 + \bar{T},
\label{bar-T}
\end{equation}
where $\bar{T}$ is a small function, while 
\begin{equation}
s = -1 +\bar{s}.
\label{bar-s}
\end{equation}
Substituting the formulas (\ref{bar-T}) and \eqref{bar-s} into Eq. (\ref{KGT}), we find that the functions $\bar{T}$ and $\bar{s}$ satisfy a simple equation 
\begin{equation}
\frac{d\bar{s}}{d\bar{T}} = \frac{\bar{s}}{\bar{T}}.
\label{corner-eq}
\end{equation}
Its general solution is 
\begin{equation}
\bar{s} = C\bar{T},
\label{corner-eq1}
\end{equation}
where $C$ is a positive constant. Remembering that $s=\dot{T}$ and choosing (for convenience) that the moment of crossing is equal to $t=0$ we can also note 
that our field crosses the corner so that 
\begin{equation}
\bar{T} = -t,
\label{corner-eq2}
\end{equation}
and 
\begin{equation}
\bar{s} = -Ct.
\label{corner-eq3}
\end{equation}
It is interesting to notice that in paper \cite{tach1} the predictions of the model, suggested in paper \cite{we-tach}, were compared with the supernovae type Ia data
and it was discovered that there were cosmological trajectories going toward the corners which were compatible with these data. 

Now, before going inside the strip to study the cosmological evolution there, let us consider what happens with particles during the transformation of the tachyon into the pseudotachyon. To do this, we add to Eq. (\ref{KGT}) the terms responsible for the contribution of the spatial derivatives  
\begin{eqnarray}
&&\frac{\ddot{T}\left(1+\frac{1}{a^2}T_{,i}T_{,i}\right)}{1-\dot{T}^2+\frac{1}{a^2}T_{,i}T_{,i}}+3\frac{\dot{a}}{a}\dot{T}+\frac{V_{,T}}{V}\nonumber \\
&&+\frac{\dot{a}a\dot{T}T_{,i}T_{,i}-2a^2\dot{T}\dot{T}_{,i}T_{,i} +T_{,i}T_{,j}T_{,ij}}{a^4\left(1-\dot{T}^2+\frac{1}{a^2}T_{,i}T_{,i}\right)}\nonumber \\
&&-\frac{1}{a^2}\Delta T=0.
\label{KG-space}
\end{eqnarray}
 Now, expressing $\frac{\dot{a}}{a}$ through the Friedmann equation 
 $$
 \frac{\dot{a}^2}{a^2} = \rho,
 $$
 and representing the tachyon field as 
 $$
 T=T_0+\tilde{T},
 $$
 where $T_0$ is the solution of the tachyon field equation for the spatially homogeneous background mode and 
 $\tilde{T}$ is the linear perturbation, we obtain the following equation for the linear perturbations
 \begin{eqnarray}
 &&\frac{\ddot{\tilde{T}}}{1-\dot{T}_0^2}+\left(\frac{2\ddot{T}_0\dot{T}_0}{(1-\dot{T}_0^2)^2}+\frac{3\sqrt{V}(2-\dot{T}_0^2)}{2(1-\dot{T}_0^2)^{5/4}}\right)\dot{\tilde{T}}\nonumber \\
 &&+\left(\frac{3V_{,T}\dot{T}_0}{2\sqrt{V}(1-\dot{T}_0^2)^{1/4}}+\frac{V_{,TT}}{V}-\frac{V_{,T}^2}{V^2}+\frac{k^2}{a^2}\right)\tilde{T}\nonumber \\
 &&=0.
 \label{linear}
 \end{eqnarray}

Then we substitute the expressions \eqref{bar-T} and \eqref{bar-s} into Eq. \eqref{linear} instead of $T_0$,  and omitting subleading terms we obtain the following differential equation for the linear perturbations
\begin{equation}
\ddot{\tilde{T}}-\frac{1}{t}\dot{\tilde{T}}+\frac{C}{t}\tilde{T} = 0.
\label{linear1}
\end{equation}
The solution is 
\begin{equation}
\tilde{T} = c_1 tJ_2\left(\sqrt{Ct}\right)+c_2 tY_2\left(\sqrt{Ct}\right),
\label{linear2}
\end{equation}
where $J$ and $Y$ are the  Bessel functions. Both solutions are regular at $t \rightarrow 0$ and 
the particles should pass through the corner. The same analysis can be carried out in the upper left corner, where the pseudotachyon field is transformed into the tachyon field while the universe is expanding. 

However, for the perturbations of the tachyon field the relations between the amplitudes of the models and their conjugate 
momenta differ from that for the minimally coupled scalar field \eqref{canon1}. Indeed, due to the nonlinearity of the Lagrangian 
\eqref{tach}, this relation looks now 
\begin{equation}
{\cal P}_{\tilde{T}} = \frac{V(T_0)}{\sqrt{1-\dot{T_0}^2}}a^3\dot{\tilde{T}}.
\label{con-tach}
\end{equation}
This means that into the Wronskian relation instead of $a^3$ one has
\begin{equation}
a^3 \rightarrow   \frac{V(T_0)}{\sqrt{1-\dot{T_0}^2}}a^3\dot{\tilde{T}}.
\label{can-tach1}
\end{equation}
Taking into account the Friedmann equation, we have
\begin{equation}
{\cal P}_{\tilde{T}} = \dot{a}^2a\dot{\tilde{T}}.
\label{con-tach2}
\end{equation} 
Correspondingly the quantum state of the vacuum is represented by the function
\begin{equation}
\Psi_0(\tilde{T}) \sim \frac{1}{\sqrt{|u|}}\exp\left(i\dot{a}^2a\frac{\dot{u}^*}{u*}\tilde{T}^2\right).
\label{wave-tach}
\end{equation}
Here, the factor $\dot{a}^2a$ is a finite number at the crossing of the corner. As follows from the formula \eqref{linear2} 
in the vicinity of the corner the basis functions behave as 
\begin{equation}
u = A + iB t^2,
\label{wave-tach1}
\end{equation}
 where $A$ and $B$ are some constant, satisfying the normalization relation. We obtain that 
 \begin{equation}
\Psi_0(\tilde{T}) \sim \exp\left(-C(-t)\tilde{T}^2\right),
\label{wave-tach2}
\end{equation}
where $C$ is a positive constant. Thus, we see that there is a difference between this formula and the formula   \eqref{Rip12},
obtained in the preceding section. Indeed, here the coefficient in front of $\tilde{T}^2$ is not a constant as it was in \eqref{Rip12}) but is proportional to $-t$. It means that at the moment of the corner crossing the Gaussian function has the infinite dispersion.
Then after the crossing at $t > 0$ it will have a form
 \begin{equation}
\Psi_0(\tilde{T}) \sim \exp\left(-Ct\tilde{T}^2\right).
\label{wave-tach3}
\end{equation}
 Thus, in this case we have a regular basis functions in the vicinity of the corner, but at the passing through it 
 the vacuum state in some manner disappear (one can interpret the infinite dispersion in this way), but immediately after the crossing we have a Fock space again. Perhaps, this momentary disappearance of the vacuum corresponds to the transformation of the particles of the tachyon field into the particles of the pseudotachyon field.

Let us remember what happens with the pseudotachyon field and the universe after the crossing the left lower corner. 
As it was described in \cite{we-tach} at some finite moment of time and  at some finite value of the tachyon field the universe encounters the Big Brake singularity, where the 
scale  factor has a finite value too, its time derivative is equal to zero, while the deceleration tends to infinity. Choosing the moment of arriving to the Big Brake as $t=0$ we can write down the expressions for the pseudotachyon field and the cosmological scale factor as follows \cite{tach2}:
\begin{equation}
T_0(t) = T_{BB} +\left(\frac{4}{3W(T_{BB})}\right)^{1/3}(-t)^{1/3},
\label{BB}
\end{equation}
\begin{equation}
a(t) = a_{BB}-\frac34a_{BB}\left(\frac{9W^2(T_{BB})}{2}\right)^{1/3}(-t)^{4/3}.
\label{BB1}
\end{equation}

Taking into account the fact the Friedmann equation  is given now by \eqref{pseudo-Fried}, the equation for the linear perturbation becomes slightly different from Eq. \eqref{linear}:
\begin{eqnarray}
&&\frac{\ddot{\tilde{T}}}{1-\dot{T}_0^2}+\left(\frac{2\ddot{T}_0\dot{T}_0}{(1-\dot{T}_0^2)^2}+\frac32\frac{\sqrt{W}(2-\dot{T}_0^2)}{(\dot{T}_0^2-1)^{5/4}}\right)\dot{\tilde{T}}\nonumber \\
&&+\left(\frac32\frac{W_{,T}\dot{T_0}}{\sqrt{W}(\dot{T}_0^2-1)^{1/4}}+\frac{W_{,TT}}{W}-\frac{W_{,T}^2}{W^2}+\frac{k^2}{a_{BB}^2}\right)\tilde{T}\nonumber \\
&&=0.
\label{p-linear1}
\end{eqnarray}   
Using the expression \eqref{BB}, we reduce Eq. \eqref{p-linear1} to the following simple form (keeping only the leading terms)
\begin{equation}
\ddot{\tilde{T}}+\frac{5}{3t}\dot{\tilde{T}}+\frac{B^2}{t^{\frac53}}\tilde{T}=0,
\label{p-linear2}
\end{equation}
where 
\begin{equation}
B^2 = -\frac{W_{,T}(T_{BB})}{16}\left(\frac{4}{3W(T_{BB})}\right)^{4/3} > 0.
\label{p-linear3}
\end{equation}
The general solution of Eq. \eqref{p-linear2} is 
\begin{equation}
\tilde{T}(t) = c_1t^{-\frac13}J_2\left(Bt^{\frac16}\right)+c_2t^{-\frac13}Y_2\left(Bt^{\frac16}\right).
\label{p-linear4}
\end{equation}
Obviously, the second term in the right-hand side of Eq. \eqref{p-linear4} is singular at $t \rightarrow 0_-$ and we cannot use two independent solutions of the differential equation \eqref{p-linear2} to construct the Fock space. Thus, when approaching the Big Brake singularity the particles in some way disappear. 

It is interesting to consider a little bit different situation when the universe encounter a more general soft singularity \cite{Paradox1}. 
Suppose that  our universe is filled not only with the tachyon field with the potential, described above \cite{we-tach}, but also with some quantity of dust.  What will happen in such universe when the energy density of the pseudotachyon field tends to zero, while its pressure tends to infinity? In this case the deceleration also tends to infinity, while the energy density of the dust is finite and, hence, the universe should continue its expansion. However, if the universe continues the expansion the energy density of the pseudotachyon field becomes imaginary. Thus, we have some kind of a paradox \cite{paradox}.    
The solution of this paradox was first found for the case of the anti-Chaplygin gas  - perfect fluid with the equation of state
$$
p = \frac{A}{\rho},\ A>0,
$$
which represents the simplest model, where the Big Brake singularity arises. The solution of the problem \cite{Paradox1} consists in the fact the equation of state of this gas undergoes a transformation and it becomes the standard Chaplygin gas, but with a negative energy density. This solution was extended to the case of the pseudotachyon, which transforms itself into the quasitachyon with the Lagrangian
\begin{equation}
L = W(T)\sqrt{\dot{T}^2+1}.
\label{quasi}
\end{equation}
Let us present in detail what happens with the pseudotachyon field when the universe in the presence of dust is running toward the future soft singularity. It behaves as
\begin{equation}
T(t) = T_s+\frac{2}{\sqrt{6H_S}}\sqrt{-t},
\label{sing-dust}
\end{equation} 
where the value of the Hubble constant at the singularity $H_S$ is found from the Friedmann equation for the universe filled with dust
\begin{equation}
H_S^2 = \frac{\rho_0}{a_{S}^3},
\label{dust}
\end{equation}
where $\rho_0$ is a positive constant.  To get the correct equation for the linearized perturbations of the pseudotachyon field in the vicinity of the singularity we use the Friedmann equation in the presence of both the pseudotachyon field and dust
\begin{equation}
\frac{\dot{a}^2}{a^2} = H_S^2 + \frac{W(T_0)}{\dot{T}^2-1}.
\label{dust1}
\end{equation}
As a result we obtain the following equation for the linear perturbations of the pseudotachyon field (where as before we keep only the leading terms in the coefficients before $\ddot{\tilde{T}}, \dot{\tilde{T}}$ and $\tilde{T}$):
\begin{equation}
\ddot{\tilde{T}}-\frac{1}{2t}\dot{\tilde{T}}+\frac{B^2}{6H_S t}\tilde{T}=0,
\label{dust2}
\end{equation}
where  
$$
B^2 = \frac{W_{,TT}(T_S)}{W(T_S)}-\frac{W_{,T}^2(T_S)}{W^2(T_S)}+\frac{k^2}{a_S^2} > 0.
$$
The solution of this equation is 
\begin{equation}
\tilde{T}(t) = c_1t^{3/4}J_{\frac32}\left(\frac{B}{\sqrt{6H_S}}t^{\frac12}\right)+c_2t^{3/4}Y_{\frac32}\left(\frac{B}{\sqrt{6H_S}}t^{\frac12}\right).
\label{dust3}
\end{equation}
Thus, we see both solutions of Eq.  \eqref{dust2} are regular, and we can construct the creation and the annihilation operators and the Fock space.  The basis functions in the vicinity of the singularity behave like 
\begin{equation}
u = D + iF(-t)^{\frac{3}{2}},
\label{dust3}
\end{equation}
and, hence, 
\begin{equation}
\frac{\dot{u}^*}{u^*} \sim \frac{iF(-t)^{\frac12}}{D}.
\label{dust4}
\end{equation}
On the other hand,  it follows from Eq. \eqref{sing-dust} that 
\begin{equation}
\frac{V(T_S)}{\sqrt{\dot{T}^2-1}} \sim \sqrt{-t}.
\label{dust5}
\end{equation}
We obtain the vacuum wave function in the form
\begin{equation}
\Psi_0(\tilde{T}) \sim \exp(-C(-t)\tilde{T}^2).
\label{dust6}
\end{equation}
We encounter the same situation which we have seen at the corner crossing: the dispersion of the Gaussian function tends to infinity at the crossing of the singularity. Nevertheless the situation looks more regular in the presence of dust. How can one interpret this fact? Perhaps, it is possible to think that the fact that the evolution at the crossing of the singularity is driven mainly by the dust makes the behaviour of the particle-like modes of the tachyon field more regular.  

\section{Phantom divide line crossing and cusped potentials}

We have already written in the section II about the phantom cosmology and the Big Rip singularity. Since the moment of the discovery of the cosmic acceleration there is a discussion about the possibility of such a cosmological evolution, where the stage of the superacceleration with $w < -1$ is a temporary one, substituted at some moment by the transition to the normal acceleration with $w > -1$. This hypothetic phenomenon is called ``phantom divide line crossing". This phenomenon can be described by models, including two scalar fields - a standard one and a phantom. More interesting option involves the consideration of the scalar field non-minimally coupled to gravity where such  effect is also possible \cite{Star-non-min0,Star-non-min}.   
In paper \cite{Vikman} rather general family of Lagrangians with the nontrivial kinetic term of the  $k$-essence type \cite{k}
was studied from the point of view of the possibility of the phantom divide line crossing. It was shown that such a phenomenon can occur, but it is unstable with respect to perturbations or the corresponding trajectories have measure zero in the space of all possible evolutions.   

 In papers \cite{cusped,cusped1} one more opportunity was considered: the cosmological evolution driven by a scalar field with a cusped potential.  Remarkably, 
a passage through the point where the  Hubble parameter  achieves a maximum value implies 
the change of the sign of the kinetic term. Though a cosmological singularity is absent in these cases, this phenomenon is a close relative of those, considered in the preceding sections, because here we also find some transformation of matter properties induced by a change of geometry. In this aspect the phenomenon of the phantom divide line crossing in the model \cite{cusped,cusped1} is  analogous to the transformation between the tachyon and pseudotachyon field in the Born-Infeld model with the trigonometric potential considered earlier. 

Consider the phantom scalar field with a negative kinetic term and the potential which has the following form
\begin{equation}
V(\phi) = \frac{V_0}{(1+V_1\phi^{\frac{2}{3}})^{2}}.
\label{cusp}
\end{equation}
The Klein-Gordon equation for the homogeneous part of the phantom scalar field has the form 
\begin{equation}
\ddot{\phi}+3\frac{\dot{a}}{a}\dot{\phi}+\frac{4V_0V_1}{3(1+V_1\phi^{\frac{2}{3}})^{3}\phi^{\frac13}}=0.
\label{cusp1}
\end{equation}
The Friedmann equation is 
\begin{equation}
\frac{\dot{a}^2}{a^2}=-\frac{\dot{\phi}^2}{2}+\frac{V_0}{(1+V_1\phi^{\frac{2}{3}})^{2}}.
\label{cusp2}
\end{equation}
We are interested in a special solution of these equations, when at some moment (we can choose it as $t=0_-$) the phantom scalar field and its time derivative tend to zero. Such a solution exists and it looks as follows
\begin{equation}
\phi(t) = \phi_0(-t)^{\frac32},
\label{cusp3}
\end{equation}
\begin{equation}
\frac{\dot{a}^2}{a^2}=\sqrt{V_0},
\label{cusp4}
\end{equation}
where 
\begin{equation}
\phi_0=\left(-\frac{16}{9}V_0V_1\right)^{\frac{3}{4}},\ V_0 > 0,\ V_1 < 0.
\label{cusp5}
\end{equation}
The analysis of the equations of motion (\ref{cusp3}) and \eqref{cusp4} shows \cite{cusped,cusped1} 
that the smooth evolution of the universe compatible with the particular initial conditions chosen in such a way to 
provide this regime is possible if at $t =0_+$ the phantom field transforms itself into the standard scalar field. This kind of the transition is indeed smooth because the kinetic term changes its sign, passing through the point when it is equal to zero. 

To explain better what happens at this passage through the point when both the field and its time derivative vanish we can recall briefly a simple mechanical analogy \cite{cusped1}. Let us consider a one-dimensional problem of a classical point
particle moving in the potential
\begin{equation}
V(x) = \frac{V_0}{(1+x^{2/3})^2},
\label{classical}
\end{equation}
where $V_0 > 0$.
The equation of motion is 
\begin{equation}
\ddot{x} -\frac{4V_0}{3(1+x^{2/3})^3 x^{1/3}} = 0.
\label{classical1}
\end{equation}
There are three types of possible motions, depending on the value of the energy $E$. If $E < V_0$, the particle cannot reach the top of the potential at the point $x=0$. If $E > V_0$, the particle passes through the top of the hill with a non-vanishing velocity. 
The case $E = V_0$ is exceptional. 
In the vicinity of the point $x = 0$ the trajectory of the particle is  
\begin{equation}
x(t) = C(t_0-t)^{3/2},
\label{classical2}
\end{equation}
where 
\begin{equation}
C = \pm \left(\frac{16V_0}{9}\right)^{3/4}
\label{C-define}
\end{equation}
and $t \leq t_0$. 
Independently of the sign of $C$ in Eq. (\ref{C-define}) 
the signs of the particle coordinate $x$ and its velocity $\dot{x}$ are opposite and hence, the particle can arrive in finite time to the point of the cusp of the potential  at $x = 0$.  
Another solution reads as 
\begin{equation}
x = C(t-t_0)^{3/2},
\label{classical4}
\end{equation}
where $t \geq t_0$.
This solution describes the particle going away from the point $x = 0$. 
Thus, we can combine the branches of the solutions (\ref{classical2}) and (\ref{classical4}) 
in four different manners and there is no way to choose if the particle arriving to the point 
$x=0$ should go back or should pass the cusp of the potential (\ref{classical}). It can stop at the top as well. 
To observe an analogy between this problem and the cosmological one we 
can try to introduce a friction term into the Newton equation (\ref{classical1})
\begin{equation}
\ddot{x} + \gamma\dot{x}-\frac{4V_0}{3(1+x^{2/3})^3 x^{1/3}} = 0.
\label{classical5}
\end{equation} 
If the friction coefficient $\gamma$ 
is a constant, one does not have 
a qualitative change with respect to the discussion above. 
However,  if  $\gamma$ is 
\begin{equation}
\gamma = 3\sqrt{\frac{\dot{x}^2}{2}+V(x)}.
\label{gamma} 
\end{equation}
then
\begin{equation}
\dot{\gamma} = -\frac{3}{2}\dot{x}^2
\label{gammadot}
\end{equation}
and 
\begin{equation}
\ddot{\gamma} = -3\ddot{x}\dot{x}
\label{gammaddot}
\end{equation}
just like in the cosmological case, where the role of the friction coefficient is played by the Hubble parameter. 
The trajectory arriving to the cusp with a vanishing velocity is still 
described by 
the solution (\ref{classical2}). Consider the particle coming to the
cusp 
from the left 
($C < 0$). It is easy to see that the value of $\dot{\gamma}$ at the 
moment $t_0$ tends to zero,
while its second derivative $\ddot{\gamma}$ given by Eq. (\ref{gammaddot}) is 
\begin{equation}
\ddot{\gamma}(t_0) = \frac98 C^2 > 0.
\label{gammaddot1}
\end{equation}
Thus, it looks like  the friction coefficient $\gamma$ reaches its
minimum 
value at $t = t_0$. 
Let us suppose that the particle is coming back to the left from
the 
cusp and its motion is described by Eq. (\ref{classical4}) with
negative $C$. 
A simple check shows that 
in this case 
\begin{equation}
\ddot{\gamma}(t_0) = -\frac98 C^2 < 0.
\label{gammaddot2}
\end{equation}
Thus, from the point of view of the subsequent evolution this point
looks 
as a maximum 
for the function $\gamma(t)$. In fact, it simply means  the second 
derivative of 
the friction coefficient has a jump at the point $t = t_0$.
It is easy to check that if instead of choosing the motion to the
left, 
we shall move forward our particle to the right from the cusp ($C>0$), 
the sign of $\ddot{\gamma}(t_0)$ remains negative as in
Eq. (\ref{gammaddot2}) 
and hence we have the jump of this second derivative again. If one
would 
like to avoid 
this jump, one should try to change the sign in Eq. (\ref{gammaddot}). 
To implement it in a self-consistent way one can substitute 
Eq. (\ref{gamma}) by 
\begin{equation}
\gamma = 3\sqrt{-\frac{\dot{x}^2}{2}+V(x)}
\label{gamma1} 
\end{equation}
and Eq. (\ref{classical5}) by 
\begin{equation}
\ddot{x} + \gamma\dot{x}+\frac{4V_0}{3(1+x^{2/3})^3 x^{1/3}} = 0.
\label{classical51}
\end{equation} 
In fact, it is exactly that what 
happens automatically in cosmology, when we change the sign of the kinetic energy term for the scalar field, 
crossing the phantom divide line. Naturally, in cosmology 
the role of $\gamma$ is played by the Hubble variable $H$.  
The jump of the second derivative of the friction coefficient $\gamma$ corresponds to the divergence of the third time derivative of the Hubble variable, which represents some kind of a very soft cosmological singularity. Thus, when we change in a smooth way the sign of the kinetic term of the scalar field, it means that whenever possible we prefer the smoothness of the spacetime geometry to the conservation of the form of the equations of motion for the matter fields.

Now, as in the preceding sections, we write down the equation for linearized perturbations of the phantom field approaching the moment of the phantom divide line crossing. Using Eqs. (\ref{cusp1}) and \eqref{cusp2}, we obtain
\begin{widetext}
\begin{eqnarray}
&&\ddot{\tilde{\phi}}+\left(3\sqrt{-\frac{\dot{\phi}^2}{2}+\frac{V_0}{(1+V_1\phi^{\frac23})^2}}
-\frac{3\dot{\phi}^2}{2\sqrt{-\frac{\dot{\phi}^2}{2}+\frac{V_0}{(1+V_1\phi^{\frac23})^2}}}\right)\dot{\tilde{\phi}}\nonumber \\
&&+\left(\frac{4V_0V_1}{9\phi^{\frac43}\left(1+V_1\phi^{\frac23}\right)^3}+\frac{8V_0V_1^2}{3\phi^{\frac23}\left(1+V_1\phi^{\frac23}\right)^4} -\frac{2V_0V_1\dot{\phi}}{\sqrt{-\frac{\dot{\phi}^2}{2}+\frac{V_0}{\left(1+V_1\phi^{\frac23}\right)^2}}\phi^{\frac13}\left(1+V_1\phi^{\frac23}\right)^3}+\frac{k^2}{a^2}
\right)\tilde{\phi}=0.
\label{cusp50}
\end{eqnarray}
\end{widetext}
Using the relations \eqref{cusp4} and \eqref{cusp5}, we reduce the previous equation to the following simple form
\begin{eqnarray}
&&\ddot{\tilde{\phi}}+3\sqrt{V_0}\dot{\tilde{\phi}}+\frac{1}{4t^2}\tilde{\phi} = 0.
\label{cusp6}
\end{eqnarray}
Here, as in all the preceding considerations we have omitted the subleading contributions to the coefficients at $\tilde{\phi}$ and 
its derivatives.
The solution of this equation in the vicinity of $t = 0$ looks as 
\begin{equation}
\tilde{\phi}(t) = c_1\sqrt{-t}+c_2\sqrt{-t}\ln(-t).
\label{cusp7}
\end{equation}
We see that both the independent solutions of Eq. \eqref{cusp7} are non-singular at $t \rightarrow 0_-$. Moreover, both of them tends to zero, while their Wronskian is constant. Thus, we can try to construct the vacuum and the Fock space. 
In the case of the minimally coupled scalar field we can directly use the formula \eqref{vac-exp5}. Note that the scale factor 
at the cusp has a finite value. Thus, all possible interesting effects are connected with the behavior of basis functions.
Let us introduce 
\begin{equation}
u = A\sqrt{-t}+iB\sqrt{-t}\ln(-t).
\label{cusp8}
\end{equation}
In this case, in the vicinity of the cusp  
\begin{equation}
\Psi_0(f) \sim \frac{1}{\sqrt{\sqrt{-t}\ln(-t)}}\exp\left(-\frac{A}{B(-t)\ln^2(-t)}f^2+\frac{i}{2t}f^2\right).
\label{cusp9}
\end{equation}
We have that at $t\rightarrow 0_-$ the dispersion of the Gaussian function tends to zero and the function becomes 
the Dirac delta function. After the crossing of the cusp the dispersion becomes regular again. One can interpret this as for a moment the vacuum and the Fock space disappear and then their reappear once again, while the particles of the phantom field become particles of the standard scalar field or vice versa.

\section{Concluding remarks}

 We have considered different scenarios of the evolution of the universe with singularities or some non-analyticities in the geometry of spacetime.  We tried to answer a simple question: is it possible to conserve some kind of notion of particle corresponding to a chosen quantum field present in the universe when the latter is approaching the singularity? For simplicity we only considered scalar fields with different types of Lagrangians.  As usual we wrote down the second order differential equations for the linear perturbations of these scalar fields and studied the asymptotic behavior of 
their solutions in the vicinity of the singularity or some other particularity of the spacetime geometry. If at least one of two independent solutions has a singular asymptotic behavior, then we cannot define the creation and the annihilation operators and construct the vacuum and the Fock space. It means that the very notion of particle loses sense. This is exactly what happens when the universe is close to the Big Bang or the Big Crunch singularity. This result looks quite natural intuitively. The situation with the Big Rip singularity, studied at the end of the second section, is little bit more involved. Considering the approach to the Big Rip singularity, we saw that the Klein-Gordon equation for a standard scalar field has two regular solutions and  we can construct explicitly the vacuum state for quantum particles as a Gaussian function of the corresponding variable.  If, instead, we consider the perturbations of phantom scalar field responsible for the super-acceleration of the universe, one of two solutions of the Klein-Gordon equation is singular, and, hence, the particles cannot be defined.  

The third section was devoted to the study of a particular cosmological model based on the tachyon field with a trigonometrical potential \cite{we-tach}. Two peculiar effects distinguish this model. First, there are transformations between different kinds of Born - Infeld type fields --tachyons, pseudotachyons and quasi-tachyons. Second, the appearance of the future Big Brake singularity or, in the presence of dust, a more general type of soft future singularity. Here, we have considered the behavior of the perturbations of the Born - Infeld type fields for three differential equations. The simplest case is the passing through the point where both the potential and the kinetic term are equal to zero. We saw that in this case both solutions of the corresponding differential equation are regular, but when passing through
the corner the vacuum state in some manner disappear (one can
interpret the infinite dispersion in this way), and immediately after the crossing we again have a Fock space. The situation is different when the universe driven by the pseudotachyon field approaches the Big Brake singularity. Here, one of the solutions is singular and the particles do not exist. Strangely, if we add to the model some quantity of dust-like matter, the character of the singularity changes slightly \cite{paradox,Paradox1}, and the differential equation for the perturbations of the pseudotachyon field has two independent regular solutions. Thus, the particles exist, and the presence of dust works as a factor ``normalizing'' the passage through the singularity. We have noticed analyzing the examples in sections II, III and IV that if a field drives the evolution toward some special points like singularities then describing the linear perturbations of this field, which serve as a tool for the definition of the vacuum state, Fock space and particles, we stumble upon singular basis functions. In the case of the model including the tachyon field and dust the evolution through the soft singularity is driven mainly by dust and not by tachyon field. That is a plausible reason for the appearance of the well-defined basis functions for the perturbations of the tachyon field. But the analysis of the vacuum wave function gives us the same situation which we saw at the corner crossing: the dispersion of the Gaussian function tends to infinity at the crossing of the singularity.   

The fourth section was devoted to the model with the scalar field with cusped potential \cite{cusped,cusped1}. Here, a particular regime exists.  If we choose the initial conditions in a special way, then the phantom scalar field can be transformed into the standard scalar field with the positive kinetic term. In other words, the phantom divide line crossing occurs. There are two regular solutions for the perturbations of the scalar field in the vicinity of the crossing point, and both of them tend to zero in the corresponding limit. The dispersion of the Gaussian function tends to zero at $t\rightarrow 0_-$ and the function becomes the Dirac delta function. After the crossing of the cusp the dispersion becomes regular again. One can interpret this as for a moment the vacuum and the Fock space disappear and then reappear once again, while the particles of the phantom field become particles of the standard scalar field or vice versa.

\section*{Acknowledgements}

The work of A.K. was partially supported by the RFBR grant No 18-52-45016. O.G. acknowledges support by the Coordena\c c\~ao de Aperfei\c{c}oamento de Pessoal de N\'\i vel Superior - Brasil (CAPES) - Finance Code 001. O.G. thanks the theoretical physics group of the Universit\`a di Bologna for its hospitality.

\end{document}